\documentclass[twocolumn,showpacs,preprintnumbers,amsmath,amssymb,superscriptaddress]{revtex4}

\usepackage{graphicx}
\usepackage{dcolumn}
\usepackage{bm}
\usepackage{braket}

\newcommand{\eye}{\mathbb I}

\begin{document}

\preprint{APS/123-QED}

\title{Multiparticle State Tomography: Hidden Differences}
\author{R. B. A. Adamson}
\affiliation{%
Centre for Quantum Information $\&$ Quantum Control and Institute
for Optical Sciences, Dept. of Physics, 60 St. George St.,
University of Toronto, Toronto, ON, Canada, M5S 1A7
}%
\author{L. K. Shalm}
\affiliation{%
Centre for Quantum Information $\&$ Quantum Control and Institute
for Optical Sciences, Dept. of Physics, 60 St. George St.,
University of Toronto, Toronto, ON, Canada, M5S 1A7
}%
\author{M. W. Mitchell}
\affiliation{ICFO - Institut de Ci\`{e}ncies Fot\`{o}niques, 08860
Castelldefels (Barcelona), Spain}
\affiliation{%
Centre for Quantum Information $\&$ Quantum Control and Institute
for Optical Sciences, Dept. of Physics, 60 St. George St.,
University of Toronto, Toronto, ON, Canada, M5S 1A7
}%

\author{A. M. Steinberg}
\affiliation{%
Centre for Quantum Information $\&$ Quantum Control and Institute
for Optical Sciences, Dept. of Physics, 60 St. George St.,
University of Toronto, Toronto, ON, Canada, M5S 1A7
}%

\date{\today}
\begin{abstract}
We address the problem of completely characterizing multi-particle
states including loss of information to unobserved degrees of
freedom.  In systems where non-classical interference plays a role,
such as linear-optics quantum gates, such information can degrade
interference in two ways, by decoherence and by distinguishing the
particles.  Distinguishing information, often the limiting factor
for quantum optical devices, is not correctly described by previous
state-reconstruction techniques, which account only for decoherence.
We extend these techniques and find that a single modified density
matrix can completely describe partially-coherent,
partially-distinguishable states.  We use this observation to
experimentally characterize two-photon polarization states in
single-mode optical fiber.
\end{abstract}

\maketitle

The development of techniques for characterizing pure and mixed
quantum states has enabled many advances in quantum information
and related fields. Whether in order to study the effects of
decoherence\cite{Whit2002}, to optimize the performance of quantum
logic gates\cite{Obri2004}, to quantify the amount of information
obtainable by various parties in quantum communications
protocols\cite{Lang2004}, or to adapt quantum error correction
protocols to real-world situations\cite{Alte2004}, it is first
necessary to obtain as complete a characterization as possible of
the state of a given quantum system (or ensemble).  In the general
case of mixed states, this involves reconstruction of a density
matrix, a mathematically complete description of the degrees of
freedom of interest in a quantum system. Entanglement with
experimentally inaccessible or `hidden' degrees of freedom
(sometimes called ``the environment'') enters the density matrix
as a reduction in the off-diagonal coherences. In some quantum
systems -- those composed of multiple particles that cannot be
individually addressed -- reduced coherence can only partially
describe the effects of hidden degrees of freedom. Another
phenomenon, distinguishability, arises when hidden degrees of
freedom provide information that could in principle be used to
tell the particles apart without necessarily leading to any
changes in the coherences of the density matrix. This paper will
show how a density matrix characterization of such states can be
performed while taking into account the decoherence and
distinguishability as distinct phenomena.

Systems of particles that cannot be individually addressed occur
commonly in quantum optics and elsewhere.  A central example is the
Hong-Ou-Mandel (HOM) effect\cite{HOM1987}, in which non-classical
interference causes photon bunching at a beamsplitter.  The effect
results in photons with the same characteristics entering the same
mode, making it impossible to individually address the photons,
i.e., to manipulate or measure them individually.  Many other major
results in the field of quantum optics such as the generation of
Bell states, the demonstration of teleportation\cite{Bouw1997},
linear optics quantum computing\cite{KLM2001}, the generation of
cluster states\cite{Walt2005} and the demonstration of quantum logic
gates\cite{Obri2003,Zhao2005} also use non-classical interference
and necessarily involve states with indistinguishable,
non-individually-addressable photons.

Numerous experiments have directly studied the properties of
non-individually-addressable
photons\cite{Bogd2004_1,Bogd2004_2,Shih1994}. The work of Bogdanov
\emph{et al.} showed that the polarization state of two photons
forms a controllable three-level system or qutrit suitable for
many protocols in quantum information\cite{Lang2004} and quantum
cryptography\cite{Kasz2003,Bech2000} and proposes a method for
performing state tomography to characterize the density matrix of
the qutrit state.  This characterization is done under the
\emph{assumption} that the two photons in the state are
indistinguishable, an assumption that is justified by measurement
of high visibility HOM-like\cite{HOM1987} two photon interference.
If that assumption were invalid the tomography method would give
an incorrect description of the state.

All photons are, of course, indistinguishable in the fundamental
sense of obeying Bose-Einstein statistics. We are concerned with
another, operational sense of ``distinguishable,'' often
encountered in discussions of multi-photon coherence.  For
example, the HOM effect acts on photons which arrive
simultaneously at a beamsplitter, but not on photons which arrive
separated by more than their coherence time. We say that the
photons could in principle be distinguished by their arrival
times, and thus do not interfere. When we refer to
distinguishability and indistinguishability in this paper it is
this kind of distinguishing information that we have in mind.
Photons can be characterized by numerous degrees of freedom
including arrival time, frequency, propagation direction,
position, transverse mode and polarization.  These degrees of
freedom may or may not be experimentally accessible, depending on
the capabilities of the experimental apparatus.  In general, we
describe as `visible' those degrees of freedom that can be
measured by a given apparatus, and as `hidden' those that cannot.
This paper will explore the effect that distinguishability in
hidden degrees of freedom due to timing information, for instance,
can have on measurements performed on a visible degree of freedom
such as polarization.

Our approach can be used on a variety of photon states such as
those created by combining non-orthogonal photon
modes\cite{Mitc2004}, those that combine the output from different
spontaneous parametric downconversion sources\cite{Bogd2004_1} and
those generated by stimulated parametric
downconversion\cite{Eise2005}.  Although this paper treats systems
of two photons, the same technique can be readily extended to a
larger number of photons. This extension will be treated in another paper\cite{Adam2006}.

\begin{table}[t]
\begin{tabular}
[c]{ccc|ccc|ccc}%
 $h$&$q$&${\bf P}$&$h$&$q$&${\bf P}$&$h$&$q$&${\bf P}$\\
\hline\hline
$0^o$&$0^o$&${\bf P}_A$&$22.5^o$&$0^o$&${\bf P}_A$&$45^o$&$0^o$&${\bf P}_A$\\
$22.5^o$&$45^o$&${\bf P}_B$&$11.25^o$&$0^o$&${\bf P}_A$&$0^o$&$22.5^o$&${\bf P}_B$\\
$45^o$&$22.5^o$&${\bf P}_A$&$22.5^o$&$0^o$&${\bf P}_B$&$22.5^o$&$22.5^o$&${\bf P}_A$\\
$0^o$&$0^o$&${\bf P}_B$%
\end{tabular}
\centering \caption{The measurement operators implemented in the
tomography experiment.  The detectors can detect either a
coincidence between two photons in the H mode thereby implementing
the projector ${\bf P}_{HH} \equiv \ket{H_1 H_2}\bra{H_1 H_2}$ or
a coincidence between the H and V modes thereby implementing ${\bf
P}_{HV} \equiv \ket{H_1 V_2}\bra{H_1 V_2}+\ket{V_1 H_2}\bra{V_1
H_2}$. A quarter- and half-waveplate at angles $q$ and $h$
respectively, placed before the detection apparatus, effectively
rotate the detection operators to $U^{\otimes 2}{\bf
P}\left(U^\dagger\right)^{\otimes 2}$ where ${\bf P}$ is either
${\bf P}_{HH}$ or ${\bf P}_{HV}$ and $U\equiv \exp
\left[i\pi\left({\bf \sigma}_z\cos{2 h}-{\bf \sigma}_x\sin{2
h}\right)\right]\exp \left[\frac{i\pi}{2}\left({\bf
\sigma}_z\cos{2 q}-{\bf \sigma}_x\sin{2 q}\right)\right]$, where
${\bf \sigma}_x$, ${\bf \sigma}_y$ and ${\bf \sigma}_z$ are the
Pauli matrices.} \label{table1}
\end{table}

If one starts from the assumption that the photons are
indistinguishable in the hidden degrees of freedom as in
\cite{Bogd2004_1} then one will use a $3\times 3$ matrix to
describe a polarization qutrit as opposed to the $4\times 4$
matrix used for distinguishable photons\cite{Jame2001}. For a more
general description one could allow for a distinguishing degree of
freedom and explore what limitations the fact that it is hidden
places on measuring density matrix elements.

A general pure state of two photons can always be written as a
superposition of tensor products of states in the visible and
hidden degrees of freedom
\begin{equation}
 \ket{\psi}=\sum_i c_i
\ket{\phi_i}_\text{vis}\ket{\chi_i}_\text{hid}, \label{thestate}
\end{equation}
where $\ket{\phi_i}_\text{vis},\ket{\chi_i}_\text{hid}$ are
eigenstates of exchange operators ${\bf X}_\text{vis}$ and ${\bf
X}_\text{hid}$ for the visible and hidden degrees of freedom,
respectively, with the same eigenvalue $\pm 1$.  The requirement
that the whole state be bosonic so that ${\bf
X}_\text{vis}\otimes{\bf X}_\text{hid}\ket{\psi} = \ket{\psi}$
guarantees that each term can be written with the visible and
hidden parts of the state either both symmetric or both
anti-symmetric. A completely general state of two photons is a
mixture of states such as $\ket{\psi}$, described by a density
matrix $\rho=\sum_j w_j \ket{\psi_j}\bra{\psi_j}$.

A reduced density matrix describing only the visible degrees of
freedom can be obtained by tracing out the hidden degrees of freedom
in $\rho$.  We define
$\rho_\text{vis}=\text{Tr}_\text{hid}\left[\rho\right]$ which is
sufficient to describe any measurement outcome on the visible
degrees of freedom.  Any such outcome has a corresponding operator
${\bf B}= {\bf B}_\text{vis}\otimes \eye_\text{hid}$ where ${\bf
B}_\text{vis}$ acts only on the visible degrees of freedom and
$\eye_\text{hid}$ is the identity operator on the hidden degrees of
freedom.  Expectation values can be written as
\begin{align}
\braket{{\bf B}}&=\text{Tr}\left[\rho \left( {\bf
B}_\text{vis}\otimes \eye_\text{hid}\right)\right]\notag\\
&=\text{Tr}_\text{vis}\left[\text{Tr}_\text{hid}\left[\rho \left(
{\bf
B}_\text{vis}\otimes \eye_\text{hid}\right)\right]\right]\notag\\
&=\text{Tr}_\text{vis}\left[\rho_\text{vis}{\bf
B}_\text{vis}\right],
\end{align}
We define projectors ${\bf P}_{S,A}$ onto the symmetric $(S)$ and
antisymmetric $(A)$ subspaces of the Hilbert space for the visible
degrees of freedom. Using ${\bf P}_S+{\bf P}_A=\eye$ and expanding
$\rho$ in terms of bosonic states as in eq.\ref{thestate} it can be
shown that $\rho_\text{vis}$ has the property that
\begin{equation}
\rho_\text{vis}={\bf P}_S \rho_\text{vis}{\bf P}_S+{\bf P}_A
\rho_\text{vis}{\bf P}_A
\end{equation}
This means that $\rho_\text{vis}$ contains no coherences between the
symmetric and antisymmetric subspaces, unlike distinguishable
particle density matrices\cite{Jame2001}. We now specialize to the
experimental situation in which the visible degree of freedom is
polarization. The symmetric space is spanned by $\left\{\ket{H_1
H_2},\ket{\psi^{+}}, \ket{V_1 V_2}\right\}$ and the antisymmetric
space by $\ket{\psi^-}$ where $\ket{\psi^{\pm}}\equiv \left(\ket{H_1
V_2}\pm\ket{V_1 H_2}\right)/\sqrt{2}$. $\rho_\text{vis}$ will be
written in this basis.

Physically the lack of coherences between symmetric and
antisymmetric states expresses lack of information about the
labeling of the photons. This is illustrated by the
distinguishable-particle states $\ket{H_1 V_2}, \ket{V_1 H_2} =
\left(\ket{\psi^+}\pm\ket{\psi^-}\right)/\sqrt{2}$, for which the
phase between $\ket{\psi^+}$ and $\ket{\psi^-}$ carries the
information about the ordering of the photons.  When this ordering
is unmeasurable the magnitude of the coherence must be zero, as in
$\rho_\text{vis}$.  Nevertheless, the populations of $\ket{\psi^-}$
and $\ket{\psi^+}$ are measurable.  For the same reason all of the
coherences between the symmetric and anti-symmetric states in
$\rho_\text{vis}$ are zero. Thus $\rho_\text{vis}$ divides naturally
into two sub-matrices, a $3 \times 3$ submatrix for the symmetric
subspace and a $1 \times 1$ submatrix for the antisymmetric subspace
as shown below:
\begin{equation}
\rho_\text{vis}=\left(\begin{array}
[c]{cc}%
\left(
\begin{array}
[c]{ccc}%
\rho_{HH,HH} & \rho_{HH,\psi^+} & \rho_{HH,VV}\\
\rho_{\psi^+,HH} & \rho_{\psi^+,\psi^+} & \rho_{\psi^+,VV}\\
\rho_{VV,HH} & \rho_{VV,\psi^+} & \rho_{VV,VV}\\
\end{array}
\right) & 0 \\

0 & \left( \rho_{\psi^{-},\psi^{-}}\right)
\end{array}
\right)
\end{equation}

If all the population is contained in the symmetric subspace then
one recovers exactly the $3\times 3$ matrix that would be measured
under the assumption of indistinguishable photons. If, however,
there is population in the antisymmetric state then the presence
of distinguishing information in the hidden degrees of freedom can
be inferred. Distinguishing information can only be detected when
it correlates to polarization. Measuring all the population to be
in $\ket{H_1 H_2}$, for instance, reveals no information about
whether the two photons are distinguishable because ordering and
polarization are uncorrelated\footnote{Explicitly, the state $\ket{H_1 H_2}_\text{vis} \ket{t_1 \tau_2}_\text{hid}$
  describes experimentally indistinguishable photons if $t=\tau$, and
  distinguishable photons otherwise.}. In such circumstances the
distinguishing information is undetectable by any apparatus that
is only sensitive to polarization, but by the same token it is
irrelevant to the outcome of any measurements made with that
apparatus.

\begin{figure}
  \centerline{
    \mbox{\includegraphics[width=\columnwidth]{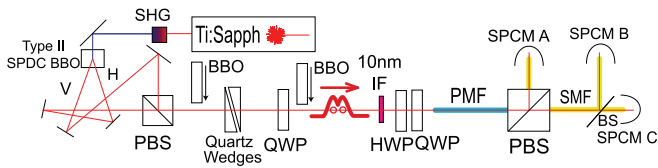}}
  }
  \caption{Experimental implementation of state preparation and tomography protocols
  showing polarizing beamsplitters (PBS), beamsplitters (BS), non-linear $\beta$-Barium
  Borate (BBO) crystals, a second harmonic generation crystal (SHG), quarter
  waveplates (QWP), half waveplates (HWP), polarization maintaining fiber (PMF), single-mode fiber
  (SMF) and single photon counting modules (SPCM).
  A spontaneous parametric downconversion (SPDC) crystal produces pairs of H and V photons. The
  separation between the H and V photons is controlled with
  movable quartz wedges and very long delays can be introduced by inserting a thick piece of BBO
  into the beam.  Single-mode fibre
  and a 10nm interference filter make the photons essentially indistinguishable
  in the spatio-temporal modes.  Tomography is performed with a set of waveplates
  and a polarizing beamsplitter.
  This system can implement all the measurements in Table \ref{table1}.}
  \label{figure1}
  \end{figure}
By tomographic measurement of the visible density matrix
$\rho_\text{vis}$ we characterize several different experimentally
produced two-photon polarization states in polarization-maintaining
single-mode fibre. As shown in Fig. $\ref{figure1}$, a 50-fs pulsed
Ti:Sapph laser centered at 810 nm was frequency-doubled to 405 nm,
pumping a spontaneous parametric downconversion crystal and creating
pairs of photons. The crystal was phase-matched in a type-II
collapsed-cone geometry\cite{Take2001} so that the photons were not
polarization-entangled, but rather emerged in separate H and V
polarized beams. These beams were recombined into a single mode on a
polarizing beamsplitter, creating a two-photon state.  A set of
matched quartz wedges was used to fine-tune the delay between the H
and V photons.  Also, a 3-mm-thick piece of BBO could be inserted
anywhere in the beam path to create a large birefringent delay. This
birefringent delay introduced distinguishing timing information that
was inaccessible to our detectors, which were not sensitive on the
100 fs timescale of the delay. This was our hidden degree of
freedom. By controlling the delay between the pulses, the degree of
overlap could be varied from completely overlapped to completely
separated. Preparation waveplates inserted before and after the
quartz wedges allowed various polarization states to be created.
Interference filters and single mode fibre served to limit
distinguishing information in the spatio-temporal mode degrees of
freedom.

The set of projectors measured in our experiment is given in Table
\ref{table1}. The detection apparatus consisted of a polarizing
beamsplitter that projected the photons to either H or V and a set
of single photon counting modules as shown in Fig.$\ref{figure1}$.
A coincidence between detectors B and C measured $\mathbf{P}_{HH}$
from Table \ref{table1}, while a coincidence between A and either
B or C measured $\mathbf{P}_{HV}$. These two fundamental
measurements were rotated with the measurement quarter-waveplate
and half-waveplate to implement the full set of measurements
listed in Table \ref{table1}.

\begin{figure}
  \centerline{
   \mbox{\includegraphics[width=\columnwidth]{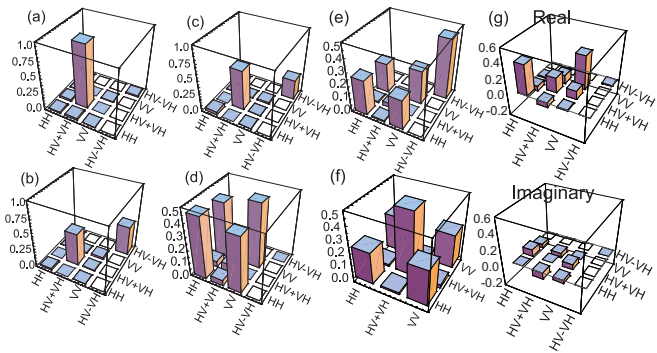}}
  }
  \caption{The density matrices measured with quantum state tomography.  The imaginary parts of (a) through (f),
  which are not shown, had all elements less than 0.05.
  White elements are inaccessible to measurement.
  (a) indistinguishable H and V photons
  (b) distinguishable H and V photons
  (c) partially distinguishable H and V photons
  (d) indistinguishable photons transformed to the 2-NOON state
  (e) distinguishable photons with the same transformation applied
  (f) the same state as it would be characterized by the technique of
  \cite{Bogd2004_2}
 (g) a 2-NOON state after passage through decohering, misaligned BBO
crystal}
  \label{figure2}
  \end{figure}

The simplest state to prepare is composed of one horizontal and one
vertical photon with a variable delay between them. When the H and V
photons are overlapped as in Fig.$\ref{figure2}a$, tomography shows
that $98\%$ of the population in the state is contained in the
completely symmetric state $\ket{\psi^+}$ indicating that the two
photons are highly indistinguishable. When the photons are delayed
by a time larger than their coherence time we obtain
Fig.$\ref{figure2}b$. The population splits with $45\%$ of the
population in $\ket{\psi^+}$ and $55\%$ of the population in
$\ket{\psi^-}$, indicating that the photons are completely
distinguishable to within the experimental limits of our
measurement.  When the delay is less than the coherence time we
obtain the state in Fig.$\ref{figure2}c$ with 62\% of the population
in $\ket{\psi^+}$ and 31\% of the population in $\ket{\psi^-}$
indicating partial hidden distinguishability.

The most widely investigated state of two indistinguishable photons
is the 2-NOON state\cite{Dowl2004,Mitc2004} which in terms of
polarization is
$\ket{2::0}_{H,V}\equiv\left(\ket{2_H,0_V}+\ket{0_H,2_V}\right)/\sqrt{2}$.
The state $\ket{1_H,1_V}$ is a 2-NOON state in the circular basis
(to within a global phase) because of the creation operator relation
${a^\dagger}_H{a^\dagger}_V=i\left({a^\dagger}_L
{a^\dagger}_L-{a^\dagger}_R{a^\dagger}_R\right)/2$. Experimentally,
a quarter waveplate can map the circular basis onto the $H/V$ basis
and turn $\ket{1_H,1_V}$ into
$\left(\ket{2_H,0_V}+\ket{0_H,2_V}\right)/\sqrt{2}$.

When the indistinguishable state in Fig.$\ref{figure2}a$ is
converted to a 2-NOON state in this way, all the population
remains confined to the symmetric subspace and the generated
state, Fig.$\ref{figure2}d$, is indeed a reasonable approximation
to the 2-NOON state with a fidelity of 0.90 and a concurrence of
$0.54$. On the other hand when the same transformation is applied
to the state with hidden distinguishability state in
Fig.$\ref{figure2}b$ the state in Fig.$\ref{figure2}e$ is
generated. The fidelity of this density matrix to the NOON state
is only 0.49, nearly identical to the 0.5 fidelity that an
incoherent mixture of HH and VV states would have to the NOON
state. Its concurrence is zero. In a multi-photon interference
experiment such as that in\cite{Mitc2004} Fig.$\ref{figure2}d$
would display high visibility interference fringes whereas the
state from Fig.$\ref{figure2}e$ would not. One might expect a
tomography protocol that assumes indistinguishable photons, such
as that proposed in \cite{Bogd2004_2}, to break down when
confronted with a state such as Fig.$\ref{figure2}d$.  To check
this we used the density matrix in Fig. $\ref{figure2}d$ to
calculate the outcomes of the measurements taken in
\cite{Bogd2004_2} and linearly reconstructed an indistinguishable
photon density matrix. The resulting matrix (Fig.
$\ref{figure2}f$) mistakenly puts all the $\ket{\psi^-}$
population in the $\ket{\psi^+}$ state.  This density matrix will incorrectly predict the outcome of measurements made in other bases such as the diagonal basis.

The state in Fig.$\ref{figure2}g$ is obtained if a 2-NOON state is
made from indistinguishable photons and then sent through a
complicated non-unitary process implemented by inserting a thick
piece of BBO whose axis is at a small angle relative to the
horizontal. As can be seen, the BBO reduces the size of the
coherence between the HH and VV terms (as well as causing some
rotation), but leaves the $\ket{\psi^-}$ term unaffected.  This
demonstrates that decoherence can occur without introducing
distinguishability between the photons.  Comparison with Fig.
$\ref{figure2}e$ shows that decoherence and distinguishability are
distinct effects with different experimental signatures.

We have developed and demonstrated experimentally state tomography
of two-photon polarization states, including, for the first time,
the effects of distinguishing information in hidden degrees of
freedom. Such distinguishing information destroys non-classical
interference, is often a limiting factor in linear-optics devices
such as quantum gates, and is not correctly described by previous
tomography schemes, which account only for decoherence.  Our
tomography technique produces a `visible density matrix' which
predicts the outcome of all polarization measurements, and which
describes the effects of both decoherence and distinguishability.
This was demonstrated clearly with our production and measurement of
a ``NOON''state affected by both decoherence and distinguishability.
The approach can be applied to other types of states and extended to
larger numbers of particles.

The authors would like to thank Peter Turner and Philip Walther for
useful discussions. This work was supported by the Natural Science
and Engineering Research Council of Canada, the DARPA QuIST program
managed by the US AFOSR (F49620-01-1-0468), Photonics Research
Ontario, the Canadian Institute for Photonics Innovation and the
Canadian Institute for Advanced Research.  MWM is supported by MEC (FIS2005-03394 and Consolider-Ingenio 2010 Project "QOIT"), AGAUR SGR2005-00189, and Marie Curie RTN "EMALI".  RBAA is funded by the Walter C. Sumner Foundation.
\bibliography{paper}
\end{document}